\documentclass[preprint,12pt]{elsarticle}

\usepackage{amssymb}
\usepackage{url}
\usepackage{xcolor}
\usepackage{xspace}
\usepackage[numbers]{natbib}
\usepackage{url}

\journal{International Journal of Medical Informatics}

\newcommand{\dabih}{\textit{dabih}\xspace}

\begin{document}
\begin{frontmatter}

\title{dabih - encrypted data storage and sharing platform} 

\author[sp]{Michael Huttner} \ead{michael.huttner@ur.de}
\author[lit]{Jakob Simeth} \ead{jakob.simeth@klinik.uni-regensburg.de}
\author[nt, ph]{Renato Liguori}
\author[nt, ph]{Fulvia Ferrazzi}
\author[sp]{Rainer Spang}

\affiliation[sp]{
  organization={Faculty of Informatics and Data Science (FIDS) University of Regensburg},
  addressline={Am Biopark 9},
  city={Regensburg},
  postcode={93053},
  state={Bavaria},
  country={Germany}
}
\affiliation[lit]{
  organization={Leibniz Institute for Immunotherapy},
  addressline={Franz-Josef-Strauß-Allee 11},
  city={Regensburg},
  postcode={93053},
  state={Bavaria},
  country={Germany}
}

\affiliation[nt]{
  organization={Department of Nephropathology, Institute of Pathology, Friedrich-Alexander-Universität Erlangen-Nürnberg},
  addressline={Krankenhausstraße 8-10},
  city={Erlangen},
  postcode={91054},
  state={Bavaria},
  country={Germany}
}
\affiliation[ph]{
  organization={Institute of Pathology, Friedrich-Alexander-Universität Erlangen-Nürnberg},
  addressline={Krankenhausstraße 8-10},
  city={Erlangen},
  postcode={91054},
  state={Bavaria},
  country={Germany}
}

\begin{abstract} 

\textit{Background:}
The secure management of sensitive clinical data, particularly human genomics data, has become a critical 
requirement in modern biomedical research. Although the necessary software and algorithms are readily 
available, their use by non-IT experts poses significant challenges. \\

\textit{Methods:}
  We developed \dabih, an open-source web application specifically designed to facilitate user-friendly encrypted 
data management. \dabih enables web-based uploading, storing, sharing, and downloading of sensitive data in any format. 
Its approach to data security involves a two-stage envelope encryption process. We combine symmetric-key encryption for data
and public-key encryption as key encapsulation mechanism. The private key necessary 
for decrypting the data remains exclusively on the owner's device. Thus, accessing data is impossible 
without explicit permission from the keyholder.

\textit{Results:}
\dabih is available open-source on GitHub
  \url{https://github.com/spang-lab/dabih}, as ready to use containers on docker
hub and includes a command line interface and a graphical bulk upload tool as
pre-built binaries. Documentation is available as part of the web application.

\textit{Conclusions:}
\dabih enables everyone to use strong cryptography for their data, while being just as simple
to use as other, non-encrypted, data storage solutions.
All the cryptography occurs seamlessly in the background as users interact with a secure web portal, simply 
by dragging and dropping files. 

\end{abstract}

\begin{graphicalabstract}
  \includegraphics[width=\linewidth]{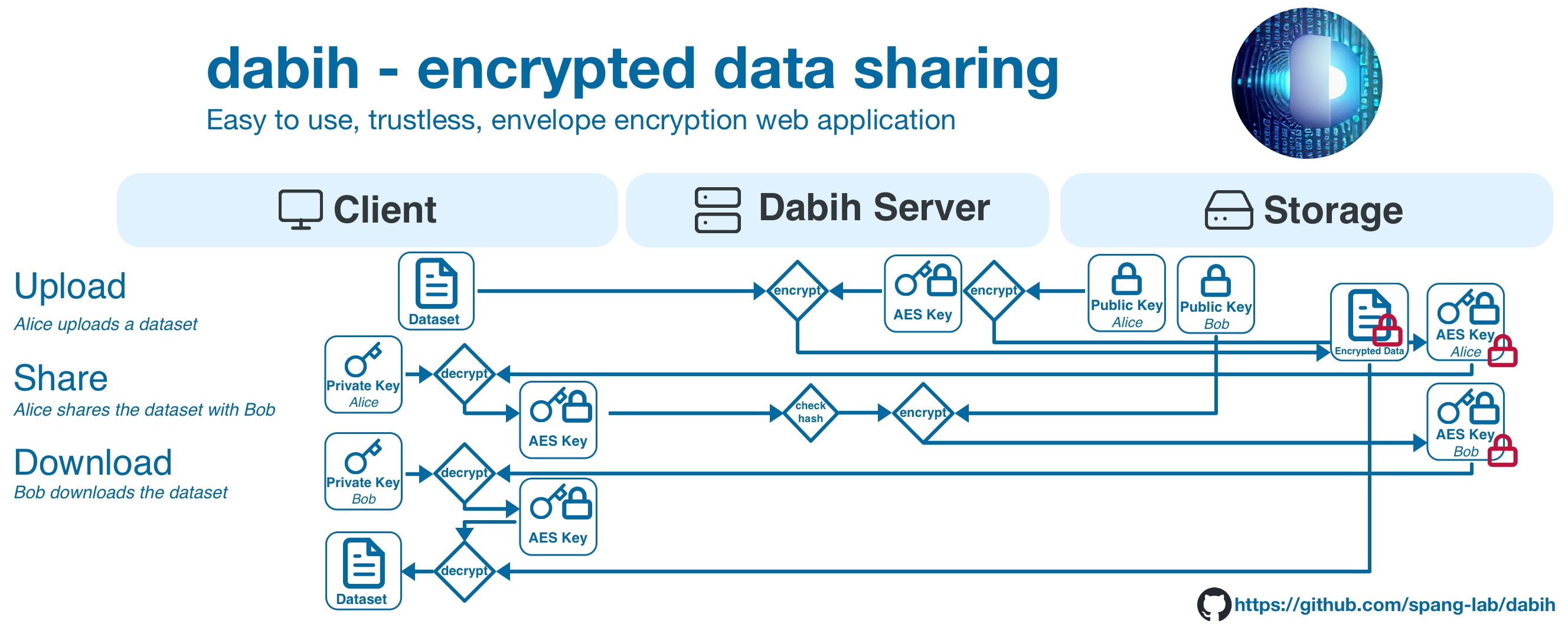}
\end{graphicalabstract}

\begin{highlights} \item Open source, self-hosted application for secure data
  sharing  \item Web application for master ease of use, API for automation, CLI
  included. \item Trustless envelope encryption security architecture. \item Users
    do not need to install software, the Web Crypto API available in web browsers is employed
  \item Easy to use, printable, QR-code based keys. \item Integration into
  OpenID based authentication systems. \item Fine-grained control and logging
  for data access. \item Cryptographic keys can be rotated. \item Options for
disaster recovery. \end{highlights}


\end{frontmatter}

\section{Introduction} 
Modern biomedical research relies heavily on large datasets, acquired by various techniques such as 
sequencing analysis or imaging.
This encompasses the acquisition, storage, sharing and analysis of highly sensitive data, including human 
genomic data. Handling such data carries significant ethical and legal implications, which are governed 
by regulations like the General Data Protection Regulation (GDPR) in the European Union.  Researchers 
must maintain stringent security measures and uphold confidentiality to protect the integrity of sensitive 
data.
For most sensitive clinical data, proper anonymization or pseudonymization are effective and practical 
solutions to protect the individual's privacy. But genomic data is special because it is identifiable 
by nature. In this case, the principle of least privilege~\cite{polp} must be rigidly 
applied. This can be achieved through the use of asymmetric encryption, 
limiting access to a minimal set of authorized individuals. Additionally, implementing fine-grained access 
control further ensures that only those authorized individuals can access the data.
 Software and algorithms for this purpose are well established, with comprehensive recommendations available, 
such as those from the German Federal Office for Information Security \cite{bsi-crypto}. 
The predominant shortcoming is the usability of these algorithms especially in integrating key management, 
authentication, and authorization. For example, the most widely used standard OpenPGP \cite{rfc4880},
implemented by the GnuPG software, requires installing software, cryptography knowledge and is built for
use in the command line.
Typically, data owners are clinicians and biomedical researchers who 
may not possess extensive IT expertise. It is crucial for them to manage their data securely while avoiding 
the complexities involved in understanding encryption and key management in detail.
To address this, here we present \dabih, an open-source web application specifically designed to facilitate 
user-friendly encrypted data management. \dabih relies on  \textit{Web Cryptography API} \cite{web-crypto}, 
a tool integrated in modern web browsers that allows us to overcome many usability and portability issues 
by employing a web application that functions within the browser.

\section{Methods}
\label{methods}

\dabih implements a hybrid cryptosystem with symmetric-key encryption for data and  public-key encryption
as key encapsulation mechanism, enabling
easy permission changes by re-encrypting only the symmetric key to authorized data recipients. The
256-bit Advanced Encryption Standard with Cipher Block Chaining (AES-256-CBC),
as specified in NIST SP800-38A\cite{nistsp800-38a} is used as the symmetric
algorithm, 4096-bit RSA (Rivest–Shamir–Adleman) with Optimal Asymmetric
Encryption Padding (OAEP) as specified in RFC3447\cite{rfc3447} is used for key
encryption.

\dabih is implemented as a server-client system.
The server (we call \textit{dabih server}) provides a web API, receives
and sends the data and manages the keys.
Clients handle the cryptography related to the private key on the users
device, e.g. the user's web browser.

\subsection{Authentication and Key management} 

\begin{figure}[h] \centering
  \includegraphics[width=0.8\textwidth]{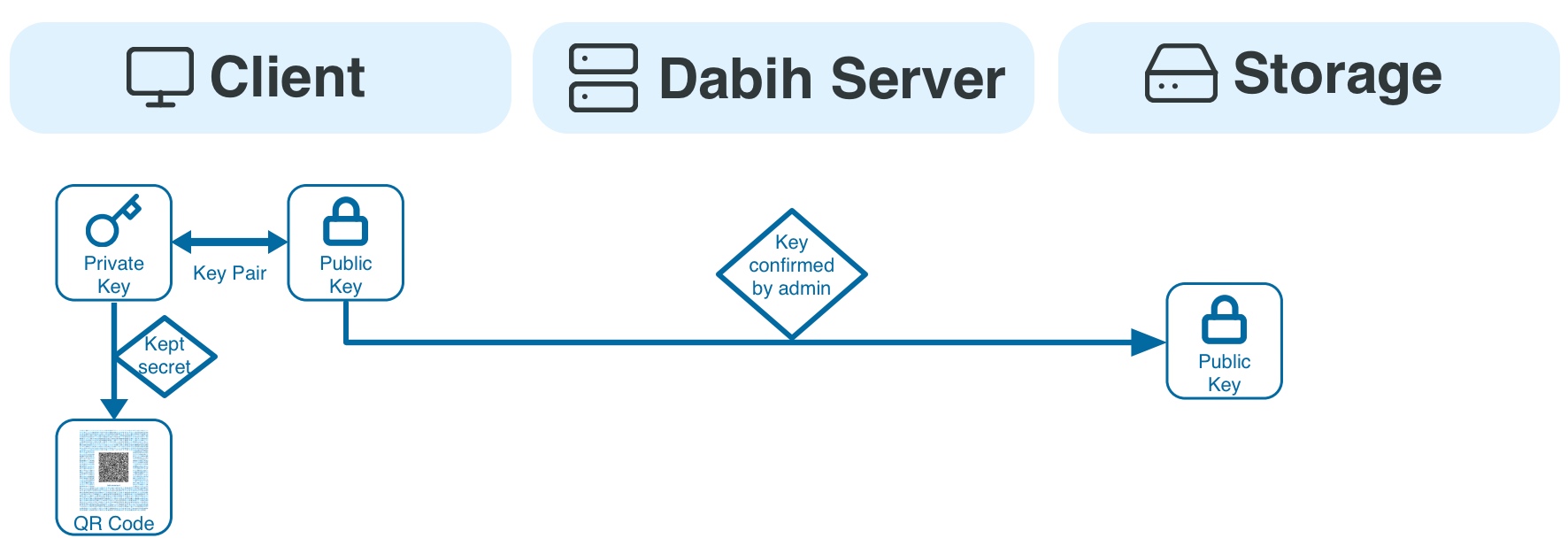} \caption{ RSA-4096
  key-pair generation and upload to \dabih. Key generation always happens locally,
  and the private key never leaves the user's computer. The public key is
uploaded to \dabih. The private key can be generated as a printable QR Code and
scanned in with a normal webcam, it will only be stored in the browsers local
storage. } \label{keyupload} \end{figure} 

For authentication, \dabih uses third-party services through the \textit{OAuth2} protocol. This 
allows server administrators to configure various authentication providers such as Google, GitHub, Keycloak, 
or any other OpenID provider, facilitating easy integration with different institutional setups. \dabih 
requests a user ID, name, and email from these providers. By default, anyone can sign up; however, to 
actively use \dabih, users must also submit an RSA public key. This key requires admin confirmation 
before it becomes active, ensuring that admins maintain control over who can access their \dabih 
instance.
The RSA key-pair can be generated by the \dabih client or by using the
popular tool \textit{ssh-keygen} \\ \texttt{ssh-keygen -m pkcs8 -t rsa -b 4096
-f key.pem} \\ \dabih requires a 4096-bit RSA key-pair in the PKCS\#8
format. 
In the following we assume users have a valid RSA key-pair and
have uploaded their public key to \dabih. We stress that the private key
never leaves the data owner's hardware and is not uploaded to the server.

\subsection{Uploading data to dabih} \label{upload}
In the context of \dabih, 'datasets' always refer to single files. For directories, clients will 
convert them into archive files prior to uploading. The upload process in \dabih begins with 
the client initiating the action by first sending
file metadata to the upload API endpoint, the server will then generate
an id and a cryptographically strong pseudo-random 32 byte AES-256-CBC key. The
id will be returned to the client, the key will be kept in memory until the
upload is complete, it is not written to disk. The AES Key is encrypted with the
RSA public key of the user uploading the data and then stored in the database.
The \dabih client will then start the data upload by splitting the file into
chunks (typically 2 MiB) and sending them to the server. For each chunk the
server will generate a random 16 byte initialization vector (IV) and encrypt the
chunk with the IV and the AES key. Only after encryption the data is written to
the storage backend, fully implementing data-at-rest encryption. \dabih also
generates a SHA-256 hash of the unencrypted data, which allows us to check and
skip duplicate files as well as to resume incomplete uploads. A checksum of the encrypted
data is calculated using the CRC-32 algorithm \cite{koopman2002crc}, this is an
emergency redundancy against data corruption, but it is important that the
storage used has adequate protection and redundancy by itself.

After all chunks have been uploaded \dabih calculates a dataset hash by concatenating
the bytes of all the chunk hashes and hashing this data again with SHA-256. We
also write a recovery file to the storage, which can be used for offline data
recovery, see \ref{offline_recovery}. See Figure
\ref{dabih_overview} a for a schematic overview on this process.

\subsection{Data sharing} 

Data owners can grant authorized recipients either \textit{read} or \textit{write} access to their datasets. 
\textit{Read} access permits downloading the data, whereas \textit{write} access provides more extensive 
permissions. With \textit{write} access, recipients can further share access with additional users, re-encrypt 
the dataset, and even delete it.

Assume \textit{User A} wants to share a dataset with \textit{User B}. 
To do so, \textit{A} downloads the encrypted AES key for the dataset. 
Locally \textit{A} decrypts the key thus obtaining an unencrypted 
AES key for the dataset. A uploads this key again --- the server itself 
does not hold a copy of it --- and
the server re-encrypts it with the public key of \textit{B}. To prevent
against key exchange attacks from \textit{A}, the server compares the
SHA-256 fingerprint of the AES key to its database and rejects the key if it
does not match. This happens for both the \textit{read} and \textit{write}
permission, the different permission levels are written to the database and
checked on API calls by the server. This process is visualized in Figure
\ref{dabih_overview} b.

\subsection{Data Download} 

Similarly to sharing data access, downloading
datasets starts by downloading and decrypting the encrypted AES key. Now the
client can simply download the encrypted data and decrypt it locally, as seen in
Figure \ref{dabih_overview} c). While this is the most secure way to download
data, users can also send the decrypted key to \dabih. The server will
then decrypt the dataset and send the raw data, offloading this computation from
the client system. Data is still encrypted in transit by the https transport
layer security (TLS) and clear-text chunks are kept in-memory.

\subsection{Data Ingestion}
\label{ingestion}
A 'side effect' of this encryption scheme is that,
technically, no private key is required for
uploading data to \dabih. This creates a way of data ingestion: users
can allow others to upload data into their account. \dabih enables this
through the use of upload tokens. Each user can create one of more access tokens
on their account page. This token can then be sent to others with a special
link. This link shows a special upload page that uses the token owners public
key for encrypting the data. No cryptographic key and no account is required to
upload. This can be very useful for securely collecting data from others or from
automatic processing pipelines, e.g. a link can be sent to a sequencing provider
to upload data directly, no other software or account required. Unfortunately
web browsers have limitations for ingesting large amounts of data, e.g. it has to
stay open during the whole upload and our code can only access files directly
selected by the user. This is why we created an optional application for data
ingestion as part of \dabih \ref{dabih_uploader}, it can be downloaded and run to upload large amounts
of data to \dabih. These upload tokens are not very secure as they are part of the URL, 
which is why they are scoped only to upload API calls, and will expire after some time
if not renewed by the user. 
  
\subsection{Key loss} \label{keyloss} 
\begin{figure}[h] \centering
  \includegraphics[width=0.8\textwidth]{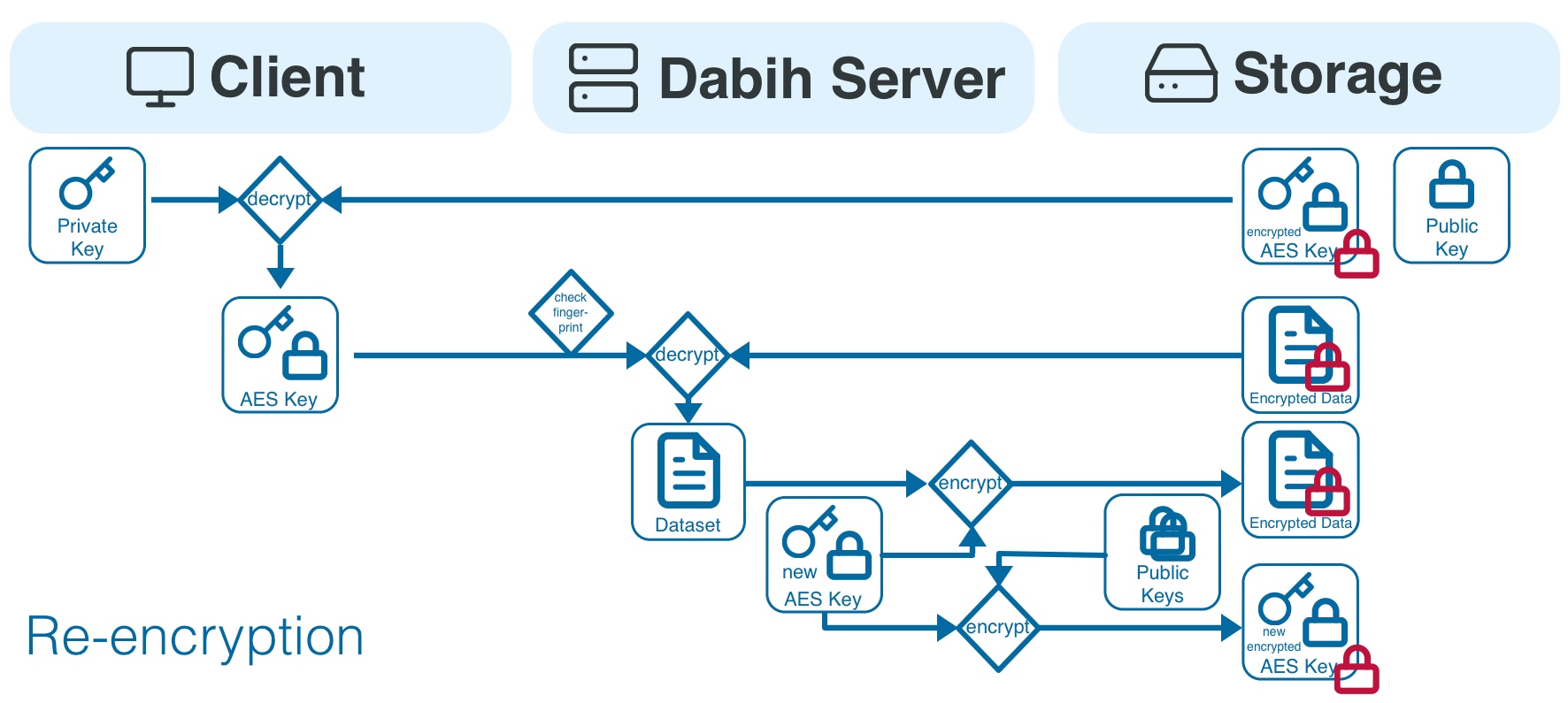} \caption{ Dataset
  re-encryption in case of key loss. As long as some other user has access to
  the data it can be re-encrypted. The user with access download the encrypted
  AES key, decrypts it and sends it to the server. The server uses the key to
  decrypt the dataset, generates a new AES key and then re-encrypts the dataset
  with the new key. All existing access permissions stay intact, the public keys
  are known to \dabih and are used to generate new encrypted AES keys. }
\label{reencryption} \end{figure}

Datasets remain encrypted throughout their entire tenure on the server, ensuring that only authorized 
data owners have the capability to decrypt them. This approach eliminates the possibility of a central 
authority, like a system administrator, recovering data if a user's private key is lost or stolen. Nevertheless, 
if at least one other user maintains access to the data, it can be re-encrypted, which allows for the 
restoration of access.
Re-encryption is done by first decrypting the AES key
for the dataset by the user who still has access to the data, or by using
a root key, see \ref{root_key}. This key is sent to \dabih, which generates a
new random AES key. All the data chunks are decrypted and then re-encrypted with
the new key. This new key is then encrypted to all the public keys of users who
are authorized to have access to the dataset. See Figure \ref{reencryption}. There is no
disruption to other users, as long as they did not store or cache the old keys
or data locally. 

\subsection{Root Keys}
\label{root_key}
If all users with access to a dataset lose their private keys, the data becomes irrecoverable. This scenario 
is particularly likely if certain datasets 
are accessible to only a single user. 
To address this issue, \dabih incorporates \textit{root keys} as an emergency backup solution, 
providing a safeguard against such situations.
Root keys are ordinary RSA-4096 key-pairs, just as every user
key. One or more public root keys can be configured for the \dabih server, and
new datasets will be automatically encrypted to every public root key. Since
root keys bypass the  security system, private keys must be stored in a
physically secure location with strict access controls and
only used for emergency recovery of the data in case all other keys are lost.
The code for recovering datasets this way is part of the \dabih command line
interface, see \ref{dabih_cli}.

\subsection{Offline Recovery}
\label{offline_recovery}
Another  disaster scenario is the loss of the
\dabih database. We stress that storage must be backed up
independently. As a precaution against loss of the database we write the most important
recovery data to disk as a part of the dataset. This recovery file contains a
list of all chunks in a dataset, with their hash, crc32 checksum and AES
initialization vector. Also included is the AES key encrypted with each public
root key. This is all the information required to decrypt the data with one of
the private root keys.

\section{Results}

\begin{figure}[h] \centering
  \includegraphics[width=0.8\textwidth]{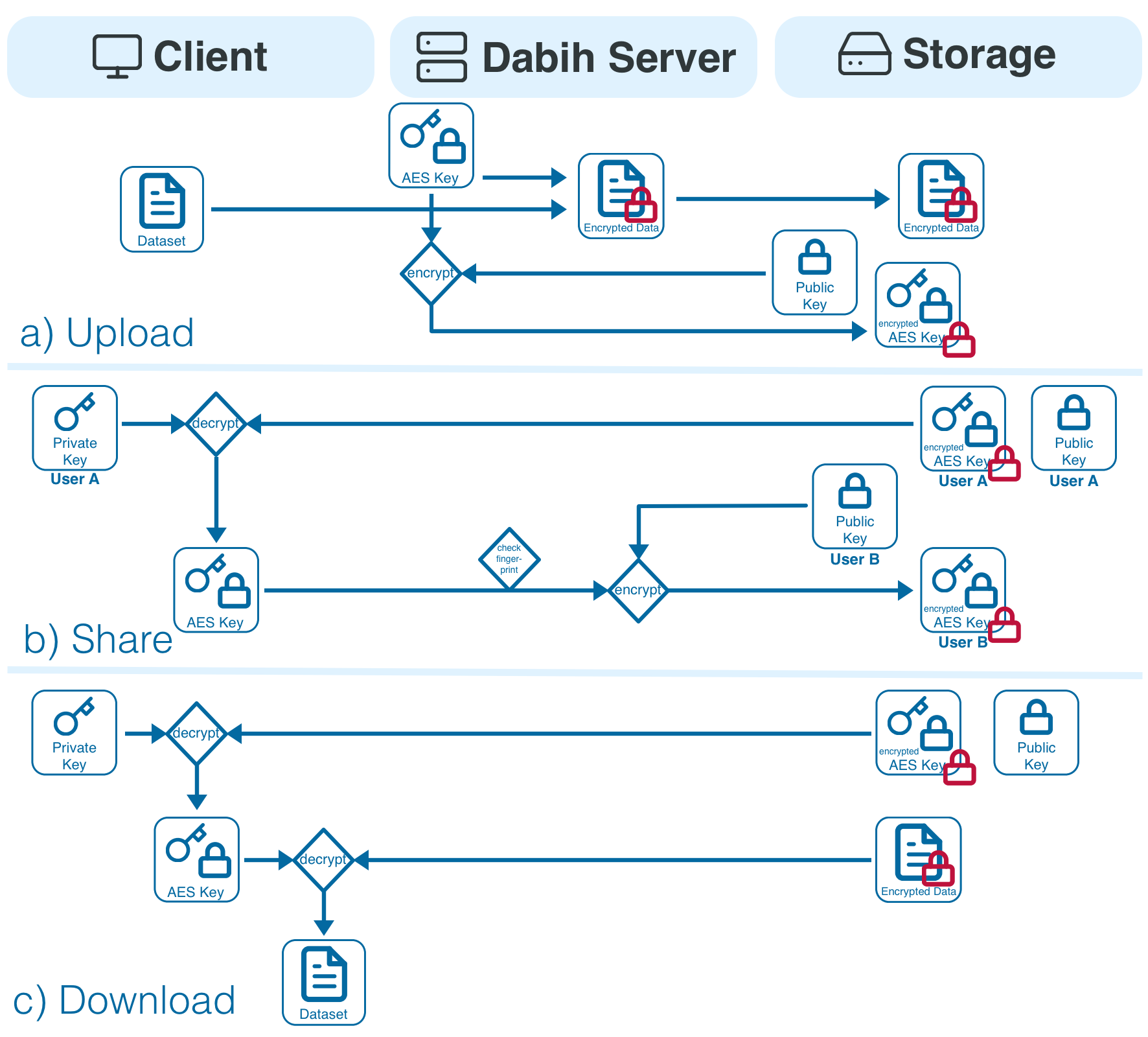} \caption{Schematic
  overview of the \dabih application. Data is
  \textbf{uploaded} to \dabih and then encrypted in two stages, the encrypted
  data is written to disk. This data can then be securely \textbf{shared} with
  authorized recipients, by re-encrypting the symmetric key with the new users public key.
  \textbf{Downloading} is the reverse of the upload process, decrypting the
data in two stages. } \label{dabih_overview}
\end{figure}

\dabih is available open-source on GitHub \cite{dabih}.
All the features described in methods are implemented by the server and web-client
in the repository. The main application flows of \textit{uploading}, \textit{sharing}
and \textit{downloading} are summarized in Figure \ref{dabih_overview}. The full code
can be audited before using \dabih to secure data. None of the security
measures rely on secrecy of the code. Documentation for deployment is provided
as part of the code repository. For ease of deployment we also provide ready to
use containers \cite{dabih-api-container}\cite{dabih-client-container} and
example deployments for different environments.
After deployment, \dabih is directly available as an easy to use
web-application. Most common administrative tasks are implemented as part of
this web application, after configuring a list of users with \textit{admin}
privileges. Administrators can enable keys, delete datasets, and access activity
logs, but they cannot access data from others. 

\subsection{Usability} While the full source code and all cryptographic details
are available, a key design consideration for \dabih is that users should
never actively need to deal with the crypto-system. After the initial setup users
just upload, share and download data. In addition to their account users only
need to manage their private key. We made this as simple as possible, allowing
users to download their key as a file or by printing it as QR Code. We encrypt
all data and do not offer a way of storing data unencrypted, as this would only
create a risk of misuse. In almost all cases the upload speed to the
\dabih server is the limiting factor for performance, the encryption
creates almost no performance cost. As \dabih primary use is in the
browser we depend on some browser API, but we only require the Web Cryptography
API and the local storage for the client, these are supported by all the major
browsers. The only notable browser we do not support is Internet Explorer,
installed on older Windows computers.

\begin{figure}[h] \centering
  \includegraphics[width=0.9\textwidth]{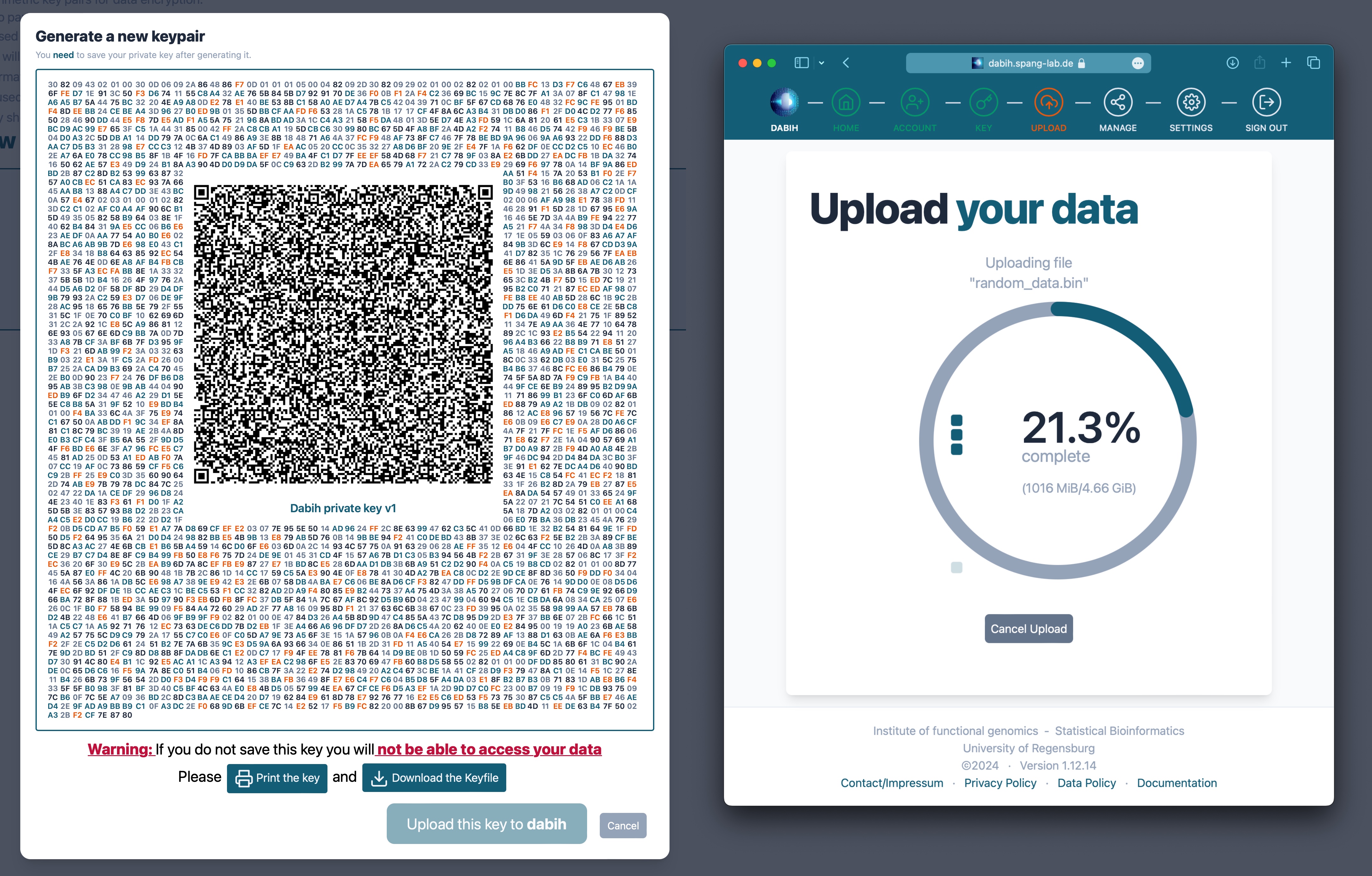} \caption{\textit{Left side:} A
  \dabih private key in a printable format. The key is encoded as a QR Code that
  can easily be read by a computer webcam. We use a special smaller format, see
  \ref{rsa_key_size}, to be able to fit the key into a single QR Code. The QR
  Code is encoded as text, to allow for easy copy and pasting of the data. The
  key is also printed out as text next to the QR Code as a redundancy measure,
  typing it in should never be required. \\
  \textit{Right side:} The dabih web client, currently uploading a large file. 
  We show a clear progress indicator, can detect duplicate uploads and can 
  resume from incomplete uploads.
} \label{privatekey} \end{figure}

  \subsection{dabih command line interface} \label{dabih_cli}

  For advanced use cases we provide an API reference as part of \dabih, after
  generating an access token \dabih can be used by any program through the web
  API. As part of the \dabih source code we also provide a \dabih command line
  interface (CLI), written in Rust. This CLI can be compiled by users or
  downloaded from our releases page on GitHub. We provide a pre-compiled binary
  for all major operating systems. It implements all the major functions of the
  graphical \dabih client but can be used in shell scripts or for other
  automation tasks. In the case of data upload we even implemented additional
  features, not possible in the browser, such as recursively searching the
  file-system, or zipping folders before upload.

  \subsection{dabih uploader} \label{dabih_uploader}

  As, by design, a private key is not required for uploading data we use this as 
  a feature for data ingress.
  A user generates a simple randomly generated upload token
  that can be sent to others and enables them to upload data to their account.
  We offer a separate app, based on the \dabih CLI \ref{dabih_cli}. This app is deliberately
  kept simple and only implements uploading to upload links for data ingestion as described in
  \ref{ingestion}. It is based on \textit{tauri}\cite{tauri} which allows us to build 
  an app for all major operating systems that calls our rust code in the \dabih CLI package. 
  It is available on our releases page on GitHub \cite{dabih}.

\section{Discussion}
\label{discussion}
  We developed \dabih as a more secure alternative to classical self-hosted file storage platforms 
  like \textit{Nextcloud} \cite{nextcloud} and \textit{Seafile} \cite{seafile}, while maintaining their 
  user-friendliness. Although these platforms provide encryption, they typically use symmetric encryption 
  with a password-derived key. \dabih enhances security by introducing asymmetric key infrastructure,. 
  Nevertheless, it is still designed for users who may not be familiar with command-line tools. Its deployment, 
  along with authentication and authorization processes, are as straightforward as those of the aforementioned 
  services. \dabih focuses exclusively on the core feature of file storage, leading to a simple and
  easy to audit system. 
  In clinical settings, \dabih can interoperate with other data management software, through it's platform
  independent and well documented API. While it is not a replacement for clinic information systems, it offers 
  a viable solution for managing sensitive research data. One of the key advantages of \dabih is 
  that it does not require custom software installations on clinic computers. Instead, all functionalities 
  are supported by standard web browsers, which are standardized and maintained by multiple large corporations. 
  The browser-based crypto-system is arguably more secure than many other libraries. This is because they 
  are widely used, ensuring that any vulnerabilities are quickly addressed. Additionally, each clinic has 
  a strong incentive to keep their employees’ browsers updated for security purposes. Differently from 
  custom data management software, which require deployment and maintenance on client devices, \dabih 
  can be provided for free, with only server maintenance needed.
  While \dabih adheres to best practices for data security, its overall security heavily relies 
  on proper user behavior. This is particularly crucial as users may not be entirely familiar with asymmetric 
  cryptography. It is imperative for data owners to maintain the confidentiality of their private keys 
  and ensure they are not lost. Additionally, \dabih's effectiveness depends upon the reliability 
  of its storage backend. The encrypted data must be regularly backed up, and the file system should prevent 
  data corruption, for instance, through the use of checksums. Due to the nature of the encryption algorithm, 
  even a single bit flip in the encrypted data can corrupt the entire file. To mitigate this, we generate 
  a checksum for every 2 MiB chunk of stored data. This is primarily for data validation and should be 
  reserved for emergency recovery purposes. 

\section{Conclusion}
\label{conclusion}

We propose \dabih as a viable and secure alternative for on-premise file storage and sharing 
solutions, while maintaining an equivalent level of user-friendliness. Leveraging the ubiquity of web 
browsers, which are installed on virtually every computer, \dabih facilitates accessible and secure cryptographic 
operations for a broad user base.

\section*{CRediT authorship contribution statement}

\textbf{Michael Huttner:} Conceptualization, Methodology, Software, Writing - Original Draft, Writing- Review \& Editing, Visualization \\
\textbf{Jakob Simeth:} Conceptualization, Methodology, Software, Writing - Review \& Editing \\
\textbf{Renato Liguori:} Validation, Writing - Review \& Editing \\
\textbf{Fulvia Ferrazzi:} Validation, Writing - Review \& Editing, Supervision \\
\textbf{Rainer Spang:} Writing - Review \& Editing, Supervision \\

\section*{Acknowledgements:}
This work was funded by the Deutsche Forschungsgemeinschaft(DFG) as part of TRR 305, project Z01.

\section*{Declaration of conflicts of interest}
All authors declare that they have no conflicts of interest.

\section*{Summary Table}

What is already known:
\begin{itemize}
    \item Protection of data is critically important for data security and user privacy.
    \item In practice, encryption and authorization are difficult to handle in a clinical setting,
      and often lead to errors or omission of required security measures.
\end{itemize}

What this study added:
\begin{itemize}
    \item By leveraging existing and wide-spread crypto software in the web browser,
      \dabih lowers the hurdle to efficient and secure encryption for non-technical end users.
    \item Cryptographic keys can be encoded and printed as a QR Code and then map closely to 
      the mental model of physical keys, familiar to users.
\end{itemize}

\bibliographystyle{elsarticle-num-names}
\bibliography{sources.bib}

\appendix

\section{Cryptography Summary}
\begin{itemize}
  \item Symmetric Algorithm: 256-bit Advanced Encryption Standard
    with Cipher Block Chaining, AES-256-CBC, specified in  NIST
  SP800-38A\cite{nistsp800-38a}.
  \item Asymmetric Algorithm: 4096 bit Rivest
    Shamir Adleman (RSA) with Optimal Asymmetric Encryption Padding (OAEP),
    specified in RFC3447\cite{rfc3447}.
  \item Hashing Algorithm: 256-bit SHA
    (SHA-256) as specified in FIPS 180-4 \cite{fips180-4}.
  \item During file
    upload the dataset is processed in memory and not written to disk.
  \item
      When the upload starts the server generates a cryptographically strong
      pseudo-random AES-256-CBC Key $k$ (24 Bytes)
    \item The client creates
      “chunks”, sequential byte buffers of the data, each with size 2 MiB.
    \item
        For each chunk we again generate a  cryptographically strong
        pseudo-random initialization vector ($iv$)
      \item The raw chunk data is
        hashed using SHA-256 and then encrypted using the AES key $k$ with the
      initialization vector $iv$
    \item We then create a crc32 checksum of the
      encrypted chunk.
    \item This encrypted chunk is then written to the file
        system and the iv, hash and checksum written to the database.
      \item All
        the asymmetric keys are RSA key-pairs, with at least 4096 bits.
      \item We
        only encrypt the 24 Byte AES Key $k$ using RSA.
      \item To prevent key
          exchange attacks all keys are fingerprinted using a SHA-256 hash.
\end{itemize}

\section{Mnemonic based id system}
\label{mnemonics}

\dabih needs a way to uniquely identify datasets,
and these identifiers will be exposed to users.
As part of the design we decided on human friendly identifiers for \dabih,
we call them \textit{mnemonics}. A \dabih mnemonic identifiers will typically
be a random adjective combined with a random first name. e.g. \texttt{vampiric\_aviyana},
\texttt{unsaluted\_esmerelda} or \texttt{branchless\_eliyana}. \dabih ensures uniqueness, 
and its name database currently contains $28476$ adjectives and $101337$ first names, allowing
for up to 2.8 billion datasets. Mnemonics have several advantages over numeric IDs, 
they are easier to remember, are simpler to exchange verbally, they prevent typing errors and 
may even provide a bit of humor.

\section{Detecting duplicate uploads}
\label{duplicate_detection}

We expect to deal with large datasets, uploading the same dataset twice can be a waste of time for 
users and tools. At the same time we always encrypt the data with different initialization vectors and 
encryption key, making it impossible to detect duplicate datasets once they are stored encrypted.
This is why we implemented a two-step hashing process during upload. We hash the original data of each 
chunk
we receive using SHA-256, and we create a dataset hash by concatenating all the chunk hashes in order 
and hashing
the hashes again. The hash behavior also is implemented separately in the CLI with the command \texttt{dabih 
hash <file-path>}.
If a client starts an upload it may send the hash of the first chunk that will be uploaded, the \dabih 
server will such a chunk
exists for the user and if it does respond with the full hash of the matching dataset. The client can
then hash the local file fully and check if it matches. If the hash matches the client can then cancel 
the upload, the default 
behavior for the currently implemented clients, but it can also continue uploading. 
This scheme allows us to detect duplicates, at almost no additional compute cost in the normal case.

\section{Restarting uploads that were interrupted.}
\label{restarting}

The hashing algorithm described in \ref{duplicate_detection} also allows us to restart interrupted uploads. 

The client may ask the server for incomplete uploads and may receive a filename, and a list of 
already complete chunks and their hashes. This allows the client to load the file again,
skipping the upload of the completed chunks, but reading and hashing them instead to ensure the data 
is identical. 
The \dabih web client and the CLI implement this feature.

\section{Reducing RSA-4096 private key size}
\label{rsa_key_size}

Our selection of cryptography algorithms is limited by what is available in the
browsers we target and by what the algorithms supported use cases are. The only
valid algorithm was RSA, with the minimum modulus length of 2048, see
\cite{nistsp800-131a}. We wanted to set the recommended modulus length of 4096,
see \cite{bsi-crypto}. At the same time we wanted users to be able to print and
scan their keys as QR Codes, as defined by ISO/IEC 18004:2015 \cite{iso18004}.
The Web Crypto API \cite{web-crypto} has functions for exporting keys. For RSA
private keys the formats  PKCS\#8 \cite{rfc5208} and JSON Web Key (JWK)
\cite{rfc7517} are supported. Unfortunately both formats are impractical for
this use case because the exported size of the key is too large. The default JWK
output is too large for any QR Code, the PKCS\#8 output produces about $2370$
bytes of output, which would only fit into a \texttt{Version 40} QR Code with
low error correction. We were unable to scan this type of code consistently with
a computer webcam. 

But the JWK Format gives us a more mathematical representation of the RSA
private key, with the following values: \texttt{n} the RSA public modulus,
\texttt{e} the RSA public exponent, \texttt{p} the smaller RSA secret prime,
\texttt{q} the larger RSA secret prime ($p < q$), \texttt{d} the RSA secret
exponent $d = e^{-1} \bmod (p-1)(q-1)$, \texttt{dp} = $d \bmod (p - 1)$,
\texttt{dq} = $d \bmod (q - 1)$, \texttt{qi} the multiplicative inverse $qi =
p^{-1} \bmod q$. 
To compress the key we can remove \texttt{n, dp, dq, qi} from the JWK and
recalculate them when we reimport the key. This results in a smaller, very
usable QR Code, even with medium level error correction, as seen in Figure
\ref{privatekey}.

\end{document}